# All-On-chip Reconfigurable Structured Light Generator

Weike Zhao[1,#], Xiaolin Yi[1,#], Jieshan Huang[2], Ruoran Liu[1], Jianwei Wang[2], Yaocheng Shi[1], Yungui Ma[1], Andrew Forbes[1,3*] and Daoxin Dai[1,*]

[1]*State Key Laboratory for Extreme Photonics and Instrumentation, College of Optical Science and Engineering, Zhejiang University, Hangzhou 310058, China.*
[2]*State Key Laboratory for Mesoscopic Physics, School of Physics, Peking University, Beijing 100871, China.*
[3]*School of Physics, University of the Witwatersrand, Johannesburg, South Africa.*

**\*** Corresponding author: dxdai@zju.edu.cn, Andrew.Forbes@wits.ac.za
[#]These authors contributed equally to this work: Weike Zhao, Xiaolin Yi

## Abstract

Structured light carrying angular momentum, such as spin angular momentum (SAM) and orbital angular momentum (OAM), has been at the core of new science and applications, driving the need for compact on-chip sources. While many static on-chip solutions have been demonstrated, as well as on-chip sources of free-space modes, no architecture that is fully reconfigurable in all angular momentum states and all on-chip has so far been possible. Here we report the first all-on-chip structured light generator for the creation of both scalar and vectorial angular momentum beams, facilitated through a silicon-on-insulator (SOI) chip with a silica mode multiplexer (silica chip). We selectively stimulate six linearly-polarized (LP) modes of the silica multimode bus waveguide, precisely controlling the modal powers and phases with the SOI chip. This allows us to tailor arbitrary superpositions of the mode set thus synthesizing common cylindrical vector vortex beams as well as OAM beams of controlled spin and topological charge. Our compact structured light generator exhibits high switching speed and operates across the telecom band, paving the way for applications such as optical communication and integrated quantum technologies.

## Introduction

In recent decades, structured light has captured a great deal of research interest and found a variety of applications[1–4], including spin angular momentum (SAM) beams[5,6], cylindrical vector (CV) beams[7,8], orbital angular momentum (OAM) beams[9,10], and total angular momentum (TAM) beams[11,12], which have special spatial distributions of intensity, phase or polarization. Among them, SAM beams are associated with circular polarization and carry SAM of S=σℏ (σ=±1) per photon, while CV beams feature a circular symmetry polarization distribution and are undetermined at the beam center where there is a polarization singularity. CV beams have shown great potential in various applications, such as plasmonic nanofocusing[13,14], particle manipulation[15], and high-resolution optical microscopy[16], due to their unique focusing property. In contrast, OAM beams have a helical phase front characterized by Hilbert factor exp($il\theta$), where $l$ is the topological charge value and $\theta$ is the azimuthal angle[17], resulting in a phase singularity at the beam center and an OAM of $l\hbar$ per photon. The unique phase/intensity distribution of OAM beams makes them very useful for widespread applications, including mode-division multiplexing[18,19], optical micro-manipulation[20–22], and optical measurement[23] to name but a few.

The myriad of applications has fuelled the generation of structured light, with the bulk optick toolkit now comprising spatial light modulators[24,25], fibre-based devices[7–9,26–30], spiral phase plates[31] and directly from lasers[32,33]. On-chip structured light generators come with the advantages of compactness, robustness, and versatility, and have attracted significant attention of late. A variety of silicon-on-insulator (SOI) photonic structures have been demonstrated for generating structured light[34–43]. On-chip metamaterials or metasurfaces imply a helical wavefront on the launched light by using a set of meta-atoms that offer a phase modulation of 0-2π with the principle of electric dipole resonance and generalized laws of reflection/refraction, and have been widely used to emit the OAM beams. By further utilizing the spin-orbit conversion, metamaterial also performs controllable transformation between SAM and OAM beams[39,40]. Alternatively, holographic gratings, which are compact and efficient, can couple an in-plane guided-mode to a free-space OAM mode by introducing subwavelength surface structures[41,42]. In addition, dielectric or plasmonic cavities with strong mode coupling have also been developed for on-chip OAM beam generation by using angular gratings to achieve free-space OAM beams with well-controlled topological charges from in degenerate whispering gallery modes (WGM). The different order cavity-modes at discrete resonance wavelengths map to the OAM beams carrying different topological charge values $l$[43], and the emitting OAM order can be tuned by using an electrical heater[38] or utilizing the strong mutual interaction of the SAM and OAM[34–37]. It should be mentioned that the aforementioned schemes are all based on the optical diffraction mechanism, and their conversion efficiency is limited to <~25% due to the downward emission to the substrate[42]. A potential solution to improve emission efficiency is coating a reflection layer on the substrate[44] or adopting unidirectional radiation grating[45]. Furthermore, the large divergence angle of diffraction structures also makes them difficult to be coupled to OAM fibres, which thus hinders their further applications.

More recently, photonic integrated circuits (PICs) have been used to create scalar structured light modes[46,47], a nascent direction that holds exciting future promise for structured light generators. In recent advances, the light was injected into the device, amplitude and phase modulated on-chip through a PIC, and then tailored in free-space by interference[47] or by using a fixed metasurface[46], producing scalar free-space modes. To realise a compact and reconfigurable approach that extends to arbitrary angular momentum beams that are controlled all on-chip (without free-space coupling) would require engineering spatial modes on the basis of on-chip waveguides or fibres and non-separably combining them with polarisation, in principle allowing all-on-chip reconfigurable angular momentum creation from scalar to vectorial states of SAM and OAM light. While this is highly desirable for fully integrated functionality, it has yet to be realised[48].

In this paper, we advance the nascent field of PICs as beam creators by demonstrating an all-on-chip reconfigurable structured light generator that produces on-demand angular momentum beams, from scalar vortex beams to vector vortex beams, all as natural fibre modes. We achieve this by using the natural modes of the silica waveguide as our basis set, controlled by six independent outputs of an SOI chip from three ports (each with two orthogonal polarisations), each reconfigurable in modal power and phase. We demonstrate this for a controlled spin and topological charge of OAM modes of order 1, as well as their vectorial combinations, including the well-known cylindrical vector vortex beams, e.g., radially and azimuthally polarised light. Our PIC architecture merges an SOI chip and a silica mode multiplexer (silica chip) for a compact device, while the all-on-chip configuration is ensured by the fibre coupled input and the subsequent on-chip excitation of scalar and vectorial OAM inside the silica multimode bus waveguide (MBW), circumventing the deleterious free-space to chip input/output coupling. Our convenient all-on-chip generation of structured light enjoys fast switching speed, a broad working bandwidth, high conversion efficiency, easy fibre-coupling (input and output), and can be extended in the future to higher OAM values and arbitrary total angular momentum by simply scaling the SOI chip output ports and MBW size without any fundamental changes to the concept or geometry.

## Results

1. **Concept and Implementation**

Our concept involves the exploitation of the SOI chip for light with controlled amplitude and phase at several output ports, using this to control the superposition of modes in a silica waveguide/fibre by edge coupling, and then finally dynamically adjusting the SOI chip to produce any desired OAM mode directly on a chip within the silica waveguide/fibre by appropriate superposition of the underlying modes. To unpack this by way of example, note that both few-mode fibres (FMFs) and square silica waveguides with appropriate cross-section parameters can support the well-known Linearly Polarised (LP) mode set, with six such LP modes (i.e., $LP_{01-x}$, $LP_{01-y}$, $LP_{11a-x}$, $LP_{11a-y}$, $LP_{11b-x}$, and $LP_{11b-y}$) shown in Fig. 1(a). From these, it is possible to construct, by appropriate superpositions, both scalar and vectorial OAM modes of topological order 1, following[3]

$$SAM_{\pm 1} = (LP_{01-x} \mp i \cdot LP_{01-y})/\sqrt{2}, \tag{1a}$$

$$RPB = (LP_{11a-x} - LP_{11b-y})/\sqrt{2} \text{ or } (LP_{11a-x} + i \cdot LP_{11b-y})/\sqrt{2}, \tag{1b}$$

$$APB = (LP_{11a-y} - LP_{11b-x})/\sqrt{2} \text{ or } (LP_{11a-y} + i \cdot LP_{11b-x})/\sqrt{2}, \tag{1c}$$

$$OAM_{\pm 1-x} = (LP_{11a-x} \pm i \cdot LP_{11b-x})/\sqrt{2}, \tag{1d}$$

$$OAM_{\pm 1-y} = (LP_{11a-y} \pm i \cdot LP_{11b-y})/\sqrt{2}. \tag{1e}$$

Here the subscript numbers are radial and azimuthal mode orders, a/b is the horizontal/vertical mode azimuth, and -x/-y is the horizontal/vertical polarisation. Higher numbers of LP modes would naturally allow higher-order OAM modes to be engineered. Continuing with our example, we see that the desired modes can be synthesized by selectively combining these six LP modes with specific modal powers and phases, which we achieve with the PIC. For instance, the $SAM_{\pm 1}$ beam can be synthesized by combining the $LP_{01-x}$ and $LP_{01-y}$ modes with a phase difference of $\pi/2$ (for right-handed circular polarization) or $-\pi/2$ (for left-handed circular polarization) according to Eq. (1a), as shown in Fig. 1(b). Note that CV beams contain the radially polarized beam (RPB) and the azimuthally polarized beam (APB), while the RPB and APB can be synthesized with the

LP$_{11a}$ and LP$_{11b}$ modes with different polarizations by following Eq. (1b) and (1c), respectively. The intensity/polarization distributions of the RPB and APB are clearly shown in Fig. 1(c), which both have donut intensity patterns, but are distinguished with radial and azimuthal polarization, respectively. Furthermore, as given in Eq. (1d) and (1e), the *x*- and *y*-polarized OAM$_{\pm 1}$ beams (i.e., OAM$_{\pm 1\text{-}x}$ and OAM$_{\pm 1\text{-}y}$) can be synthesized by combining the LP$_{11a}$ and LP$_{11b}$ modes with the same polarization. Fig. 1(d) shows the doughnut-shaped field distributions of the OAM$_{\pm 1\text{-}x}$ and OAM$_{\pm 1\text{-}y}$ beams as well as the spiral phase structures for their E$_x$ and E$_y$ components.

As a summary, the aforementioned structured light can intuitively be synthesized by exciting two specific power-, polarization- and phased-controlled LP modes. Even though LP modes are well supported in low-index-contrast optical waveguides (such as silica waveguides or fibres), the manipulation of the polarization state and the phase shifting becomes very inconvenient due to the low polarization-selectivity and low thermal-tuning efficiency. In contrast, when using silicon waveguides (which have a high index contrast), it becomes very flexible to manipulate the polarization states, the power ratios, and the phase shifts of the guided modes. As a consequence, here we propose a new PIC structured light generator by incorporating an SOI chip and a silica chip, as shown in Fig. 1(e). In particular, the SOI chip is used for manipulating the polarization states, the power ratios and the phase-shiftings of the guided light, so that three pairs of TE$_0$/TM$_0$ modes are generated with flexibly tunable power ratios and phase shifts. The silica chip is used to receive these three pairs of TE$_0$/TM$_0$ modes from the SOI chip and then multiplex them to the six quasi LP modes basis sets (i.e. LP$_{01\text{-}x}$, LP$_{01\text{-}y}$, LP$_{11a\text{-}x}$, LP$_{11a\text{-}y}$, LP$_{11b\text{-}x}$, LP$_{11b\text{-}y}$) supported in a silica MBW. With such a configuration, the power ratios and phase shifts of these six LP modes can be controlled freely by tuning the heaters on the SOI chip, and thus the SAM$_{\pm 1}$, OAM$_{\pm 1\text{-}x}$, and OAM$_{\pm 1\text{-}y}$ and RPB/APB beams can selectively be synthesized.

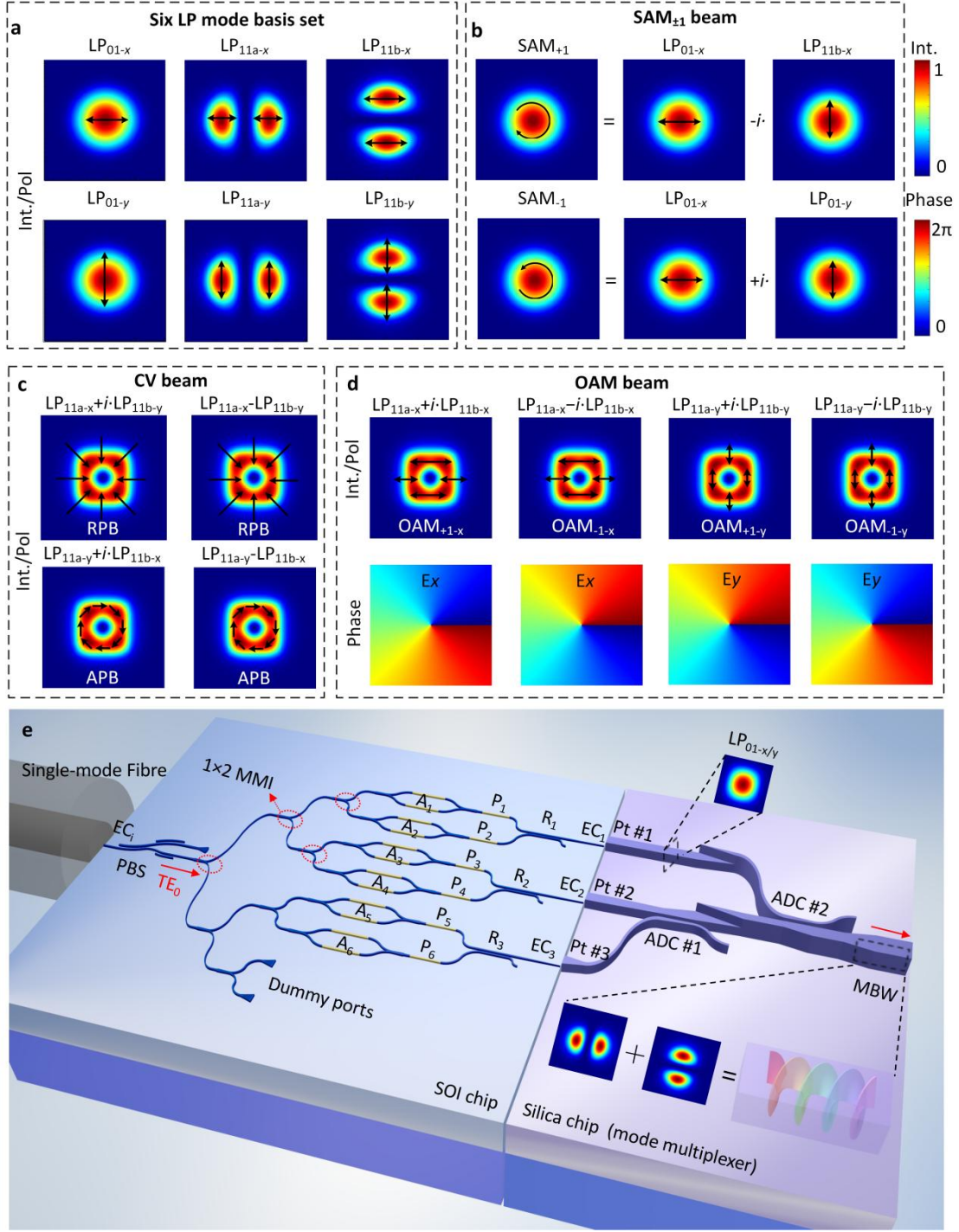

**Fig. 1 All-on-chip structured light generator.** (a) Six LP mode basis sets at a few-mode fibre or a square silica waveguide, i.e., $LP_{01-x}$, $LP_{01-y}$, $LP_{11a-x}$, $LP_{11a-y}$, $LP_{11b-x}$, and $LP_{11b-y}$ modes. (b) Top panel: the $SAM_{+1}$ beam (right-handed circular polarization) is synthesized by the $LP_{01-x}$ and $LP_{01-y}$ modes with a phase difference of $\pi/2$; bottom panel: the $SAM_{-1}$ beam (left-handed circular polarization) is synthesized by the $LP_{01-x}$ and $LP_{01-y}$ modes with a phase difference of $-\pi/2$. (c) The intensity/polarization patterns of the CV beams containing RPB and APB, which both have donut intensity patterns, but are distinguished with radial and azimuthal polarization, respectively. (d) The intensity/polarization patterns and phase structures of the $OAM_{\pm1-x}$ and $OAM_{\pm1-y}$ beams. (e) Schematic configuration of the proposed all-on-chip structured light generator incorporating an

SOI chip and a silica chip. Upper inset: the $LP_{01-x/y}$ mode in a silica single mode waveguide; Bottom inset: the $OAM_{\pm1}$ beam is synthesized as an example by using the $LP_{11a}$ and $LP_{11b}$ modes in a silica multimode bus waveguide (MBW). Operating principle: the light emitted by the fiber to the SOI chip is polarized with a polarization beam splitter (PBS) and then split into six $TE_0$ modes with a series of 1×2 multimode-interference (MMI) 3-dB couplers. These six $TE_0$ modes respectively pass through a variable optical attenuator (VOA) and a phase shifter (PS), and then are combined into three pairs of $TE_0/TM_0$ modes by three polarization splitter-rotators (PSR, $R_1$-$R_3$). These three pairs of $TE_0/TM_0$ mode are coupled to the $LP_{01-x}$/ $LP_{01-y}$ modes of three silica waveguide ports (Pt #1, Pt #2, and Pt #3) by three edge couplers (ECs, $EC_1$-$EC_3$). Finally, these three pairs of $LP_{01-x}/LP_{01-y}$ modes stimulate the six LP mode basis sets of the MBW through a silica mode multiplexer (containing two adiabatic directional couplers, ADC #1 and ADC #2). One can selectively synthesize the $SAM_{\pm1}$, $OAM_{\pm1-x}/OAM_{\pm1-y}$ beams, as well as the CV beams by controlling the powers and phases of these six LP modes with the six VOAs ($A_1$-$A_6$) and PSs ($P_1$-$P_6$).

As shown in Fig. 2(a), light from a single-mode fibre is coupled to the input port of the SOI chip with the assistance of silicon edge couplers ($EC_i$) and is polarized to be the $TE_0$ mode by using a polarization beam splitter (PBS)[49]. The $TE_0$ mode is then split by using a 1×8 power splitter based on 1×2 multimode-inteference (MMI) 3-dB couplers in cascade. Note that there are two dummy ports used for test monitoring. For these six $TE_0$ modes, there are six variable optical attenuators (VOAs; $A_1$, $A_2$, …, $A_6$) to manipulate the power ratios and six phase shifters (PSs; $P_1$, $P_2$, …, $P_6$) to manipulate the phase shifts[50], respectively. These six $TE_0$ modes with the target power ratios and phase shifts are then recombined to be three $TE_0/TM_0$ mode-pairs by using three polarization splitter-rotators (PSRs; $R_1$, $R_2$, $R_3$)[51,52] and output finally from Ports (Pt) #1, #2, and #3, respectively (see Methods). The SOI chip is butt coupled to the silica chip with three ECs ($EC_1$, $EC_2$, $EC_3$), in which way the three pairs of $TE_0/TM_0$ modes in the three silicon waveguides are coupled respectively to the $LP_{01-x}/LP_{01-y}$ modes of three silica single mode waveguides. Finally, these $LP_{01-x}/LP_{01-y}$ modes are converted and multiplexed to the six LP modes basis sets (i.e. $LP_{01-x}$, $LP_{01-y}$, $LP_{11a-x}$, $LP_{11a-y}$, $LP_{11b-x}$, $LP_{11b-y}$) supported in the MBW, by using a polarization-insensitive silica mode multiplexer, which consists of two adiabatic directional couplers (ADCs; ADC #1, ADC #2) in cascade. In detail, the $LP_{01}$ mode launched from Port #1 couples to the $LP_{11a}$ mode of the MBW via ADC #2, while the $LP_{01}$ mode launched from Port #3 couples to the $LP_{11a}$ mode of the MBW via ADC #1 and then is rotated to the $LP_{11b}$ mode with a mode rotator based on a dual-layer silica waveguide. Here the $LP_{01}$ mode launched from Port #2 transmits through these two ADCs directly. The transmission of the whole PIC structured light generator can be depicted by the transmission matrix method (see Methods). Note that while we have outlined this concept based on the implementation of our silica MBW supporting just six modes, it can be generalised to an arbitrary mode number for both higher OAM orders and TAM control.

2. **Fabrication and measurement results**

The SOI and silica chips were fabricated separately with their own regular processes. The PBSs and PSRs fabricated on the SOI chip all have a low loss of <1 dB and crosstalk <-15 dB for both $TE_0$ and $TM_0$ mode channels (see Methods). The input fibre array, SOI chip, silica chip and FMF were butt-coupled, as shown in Fig. 2(a). Here an ultra-high-NA single mode fibre (HSMF) with a numerical aperture of 0.41 and a core diameter of ~2.4 μm was used for the butt coupling with the SOI chip through an EC, enabling a relatively low coupling loss of ~2 dB for the $TE_0$ mode in the wavelength range of 1520-1600 nm. The coupling loss between the silicon waveguide and the 4×4 μm$^2$ single mode silica waveguide with the help of the same EC is 0.4/1.5 dB for the $TE_0/TM_0$ modes (see Methods). The silica mode multiplexer has a low on-chip loss of < 0.8 dB and low inter-mode crosstalk of <-14.2 dB for all three modes in the broad wavelength range of 1530-1620 nm[53]. The SOI chip was wire-bonded onto a printed circuit board (PCB) for electrical control. Fig. 2(b) shows the enlarged views of the SOI and silica chips, and the silica waveguide is labeled with a dotted line. The scanning electron microscope (SEM) images of the PBS, MMI coupler, PSR, and silica waveguide are shown in Fig. 2(c)-(f), respectively.

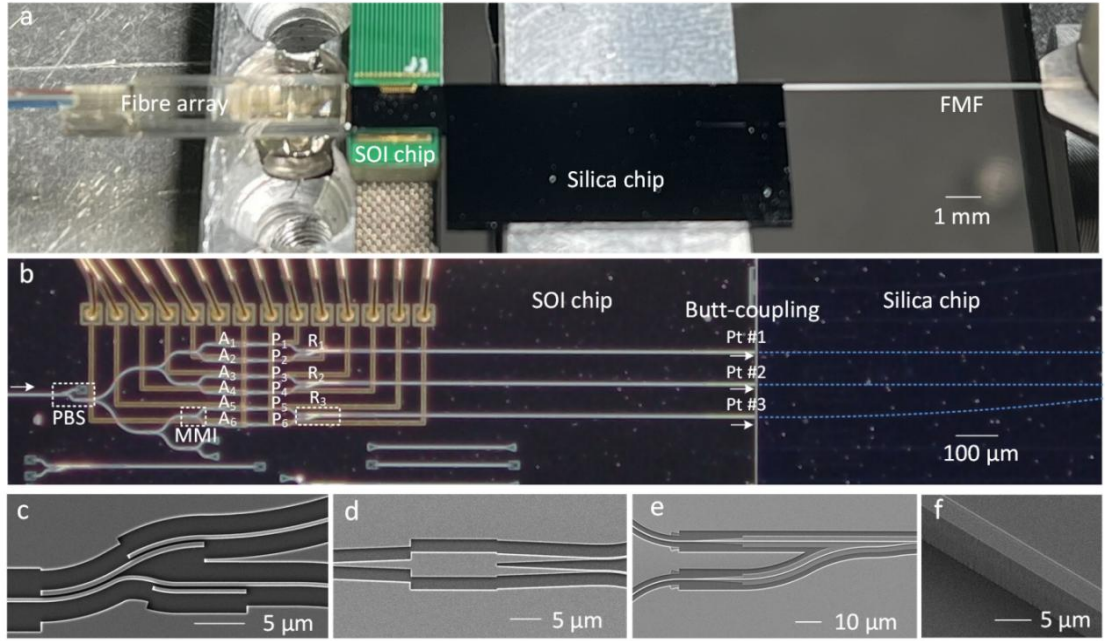

**Fig. 2** **Images of the fabricated device.** (**a**) The butt-coupled input fibre array, SOI chip, silica chip and FMF. (**b**) The enlarged view of the butt-coupled SOI chip and silica chip, where the PBS, MMI, VOAs ($A_1$-$A_6$), PSs ($P_1$-$P_6$), PSRs ($R_1$-$R_3$), three ports (Pt #1-#3) are labeled. Scanning electron microscope (SEM) images were given for the (**c**) PBS, (**d**) MMI coupler, (**e**) PSR, and (**f**) silica waveguide.

### 3. Generation of the six LP-mode basis sets

A multichannel voltage source (MVS) was used to power the six VOAs and six PSs, so that the power ratios and the phase shifts of these six $TE_0/TM_0$ channels can be controlled. Fig. 3(a)-3(b) gives the measurement results for the VOAs, showing an excess loss of < 0.6 dB and a maximum extinction ratio of ~26 dB at the wavelength of ~1580 nm. The thermal tuning of the VOAs and PSs works with a rise time $T_r$ of 11.6 μs and a falling time of $T_f$ of 7.1 μs, enabling fast generation of structured light beams. The amplitude matrix of six LP mode basis sets is given as **A**=[$A_1$, $A_2$, $A_3$, $A_4$, $A_5$, $A_6$], where $A_m$ is the amplitude for the *m*-th mode-channel, and one has $A_m$=1 or 0 by unheating or heating the *m*-th VOA ($VOA_m$). As a result, appropriately controlling the VOAs enables to individually excite any one of the $LP_{01-x}$, $LP_{01-y}$, $LP_{11a-x}$, $LP_{11a-y}$, $LP_{11b-x}$, and $LP_{11b-y}$ modes in the MBW, as shown in Fig. 3(c) (see Methods). It can be seen that all six LP modes are generated successfully with the desired mode profiles, and the perfect azimuth orthogonality of the $LP_{11a}$ and $LP_{11b}$ modes verifies that the fabricated silica chip performs well. Fig. 3(d) shows the measured transmissions from the input fiber to the output FMF when any one of these six LP modes is excited individually. It can be seen that the SOI and silica chips for the present structured light generator work well with a low excess loss of about 4-8 dB in a broad wavelength range of 1530-1595 nm (see Methods). Here the excess loss contains the total on-chip losses of about 2 dB, the coupling loss of 0.4/1.5 dB between the silicon and silica chips for x-/y- polarization, and the coupling loss of ~2-4 dB between the silica chip and the few-mode fibre. It is possible to reduce the excess loss further by improving the fabrication.

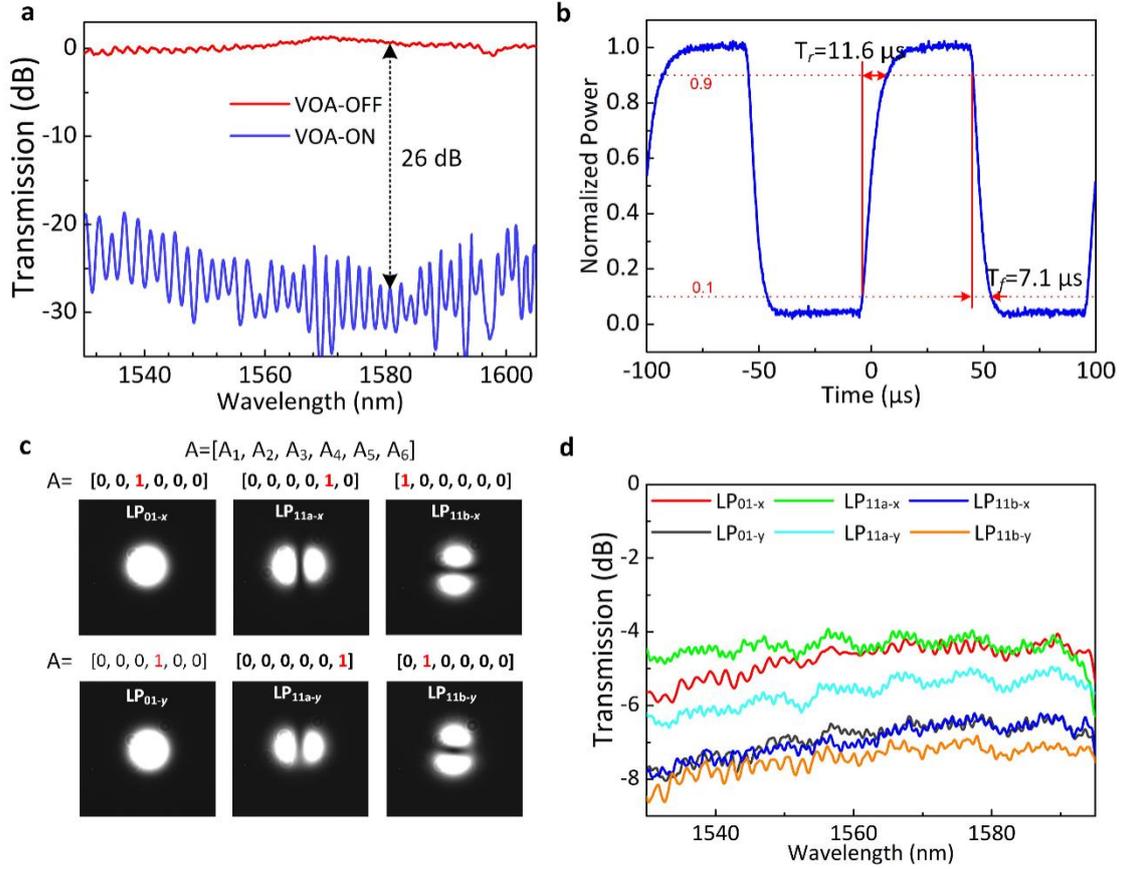

**Fig. 3** **Characterization of VOAs and the excitation of the six LP mode basis sets. (a)** The measured on/off transmissions of the testing VOA fabricated on the same SOI chip, showed a maximum extinction ratio of ~ 26 dB at 1580 nm. **(b)** The measured temporal response of the VOA, exhibits a rise time $T_r$ of 11.6 μs and a falling time $T_f$ of 7.1 μs. **(c)** Any one of six LP-modes excited individually in the MBW by controlling the VOAs to achieve the target amplitude matrix A=[$A_1$, $A_2$, $A_3$, $A_4$, $A_5$, $A_6$] as desired. **(d)** The measured transmissions from the input fiber to the output FMF when any one of these six LP modes is excited individually, show an excess loss of 4-8 dB for all six modes in a broad wavelength range of 1530-1595 nm.

## 4. Synthesis of SAM

For the generation of the $SAM_{\pm1}$ beam, the matrix **A** was set to [0, 0, 1, 1, 0, 0] so that only the $LP_{01-x}$ and $LP_{01-y}$ modes in the MBW were excited. The modal powers of these two LP modes were balanced by tuning VOA #3 and #4, while their phase difference was tuned to be $\pi/2$ or $-\pi/2$ according to Eq. (1a) by adjusting the bias voltages $V_{P3}$ and $V_{P4}$ applied to the phase shifters $P_3$ and $P_4$. The generated $SAM_{\pm1}$ beam was characterized by using a common quarter-wave plate (QWP) rotation method (see Methods), as shown in Fig. 4(a). Here the phase difference becomes $\pi/2$ and $-\pi/2$ when $V_{P3}$=1.95 V and 3.5 V, respectively. When $V_{P3}$=1.95 V, the mode intensity remained unchanged as the polarizer was rotated with different orientation angles of 45°, 90°, and 135° and 180° (see the black arrows) if no QWP was inserted. In contrast, when a QWP with a horizontal fast axis was placed in front of the polarizer, one observed constructive and destructive interference when the polarizer was 45°- and 135°-orientated, respectively, as shown in the middle panel of Fig. 4(a), indicating that the generated beam carries $SAM_{+1}$ as expected. When further increasing $V_{P3}$ to 3.5 V, constructive and destructive interference respectively happens when the polarizer is 135°- and 45°-orientated, as shown in the bottom panel of Fig. 3(a), indicating the $SAM_{-1}$ beam is generated as desired. Note that the polarization state of the output beam can traverse the whole Poincaré sphere by setting $A_2$, $A_3$ and $V_{P2/3}$ appropriately.

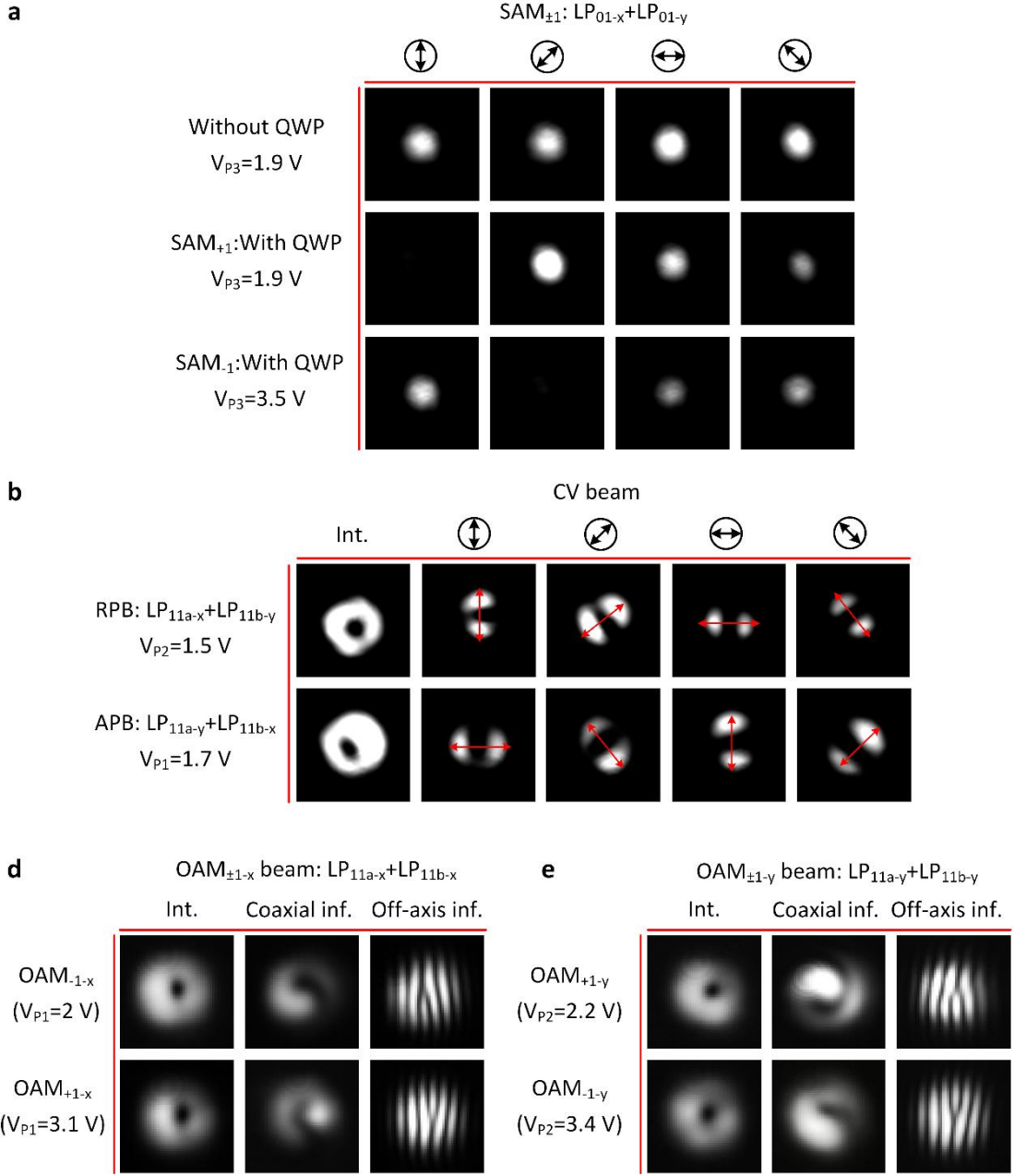

**Fig. 4 The experimental synthesis of structured light beams.** (a) The $SAM_{\pm1}$ beam is synthesized by incorporating the $LP_{01-x}$ and $LP_{01-y}$ modes and characterized with the quarter-wave plate (QWP) rotation method. Top panel: the synthesized structured light ($V_{P3}$=1.9 V) passes through a polarizer with different orientations (see the black arrows), and the mode intensity remains unchanged; Middle panel: after inserting a QWP behead the polarizer, constructive interference occurs at the 45°-orientated polarizer, and destructive interference occurs at the 135°-orientated polarizer, confirming the structured light carries $SAM_{+1}$; Bottom panel: when further increasing $V_{P3}$ to 3.5 V, the constructive interference and the destructive interference exchange with each other, confirming the structured light carries $SAM_{-1}$. (b) The CV beams are synthesized by incorporating the $LP_{11a}$ and $LP_{11b}$ modes with different polarizations. Top panel: the RPB is synthesized with the combination of $LP_{11a-x}$ and $LP_{11b-y}$ modes at $V_{P2}$=1.5 V. It has a donut intensity pattern, and maps to an $LP_{11}$ mode with a mode azimuth (yellow arrow) same as the orientation of the polarizer; Bottom panel: the APB is synthesized with the combination of

LP$_{11a-y}$ and LP$_{11b-x}$ modes at V$_{P1}$=1.7 V, and it also has a donut intensity pattern, but maps to an LP$_{11}$ mode with a mode azimuth angle vertical to the orientation of the polarizer. (**c**) The OAM$_{\pm 1-x}$ beam is synthesized with the combination of the LP$_{11a-x}$ and LP$_{11b-x}$ modes. The intensity, coaxial interference (inf.), and off-axis interference patterns confirm the generation of OAM$_{-1-x}$ (top panel, at V$_{P1}$=2 V) and OAM$_{+1-x}$ (bottom panel, at V$_{P1}$=3.1 V). (**d**) The OAM$_{\pm 1-y}$ beam is synthesized with the combination of the LP$_{11a-y}$ and LP$_{11b-y}$ modes. The intensity, coaxial interference (inf.), and off-axis interference pattern confirm the generation of OAM$_{-1-x}$ (top panel, at V$_{P2}$=2.2 V) and OAM$_{+1-x}$ (bottom panel, at V$_{P2}$=3.4 V). (@1550 nm)

## 5. Synthesis of CV beams

When matrix **A** is set to be [0, 1, 0, 0, 1, 0], the LP$_{11a-x}$ and LP$_{11b-y}$ modes in the MBW are excited simultaneously, and their phase difference can be tuned to 0 or -π/2, in which case the RPB beam is generated according to Eq. (1b). As shown by the top panel in Fig. 4(b), the synthesized mode field has a donut-shaped intensity when V$_{P2}$=1.5 V. When a polarizer with rotated orientation angles is inserted before the CCD camera (see Methods), a series of LP$_{11}$ mode fields with different azimuth are achieved (yellow arrow) and the LP$_{11}$ mode's azimuth is perfectly consistent with the polarizer orientation angles, verifying that the generated CV beam is RPB. In contrast, matrix **A** should be set as [1, 0, 0, 0, 0, 1] to synthesize the APB according to Eq. (1c), in which way the LP$_{11a-y}$ and LP$_{11b-x}$ modes are excited simultaneously. A donut intensity pattern appears when V$_{P1}$=1.65 V, as shown in the bottom panel of Fig. 4(b), while the azimuth of the LP$_{11}$ mode is perpendicular to the polarizer's transmission axis, as expected.

## 6. Synthesis of OAM beams

To generate OAM$_{\pm 1}$ beams, the LP$_{11a}$ and LP$_{11b}$ modes with the same polarization should be stimulated according to Eqs. (1d) and (1e). To verify the phase information of the generated OAM beams, an interference measurement method was used (see Methods). For the synthesis of OAM$_{+1-x}$, the LP$_{11a-x}$ and LP$_{11b-x}$ modes should be excited simultaneously and accordingly matrix **A** is set as [1, 0, 0, 0, 1, 0]. As shown in the top panel of Fig. 4(c), one clearly observes a donut-shaped intensity pattern, a clockwise helical coaxial interference pattern, and a forked off-axis interference pattern when setting V$_{P1}$=2.0 V for achieving the desired π/2 phase difference between the LP$_{11a-x}$ and LP$_{11b-x}$ modes, which indicates that the OAM$_{-1}$ beam is synthesized successfully. By further increasing the applied voltage V$_{P1}$ to 3.1 V thus adjusting the phase difference to -π/2, one can also generate the OAM$_{+1-x}$ beam, as shown in the bottom panel of Fig.

4(c). It can be seen that the generated OAM$_{+1-x}$ beam has a donut-shaped intensity pattern, an anti-clockwise helical coaxial interference pattern, and an inverse-forked off-axis interference pattern. Similarly, OAM$_{+1-y}$ and OAM$_{-1-y}$ beams can also be synthesized with the LP$_{11a-y}$ and LP$_{11b-y}$ modes by making the phase difference be -π/2 and π/2 when setting V$_{P2}$=2.2 and 3.1 V, respectively, as shown in Fig. 4(d).

## 7. Extension to TAM

Theoretically, the proposed all-on-chip structured light generator can synthesize a TAM beam that carries both OAM and SAM. A TAM beam is usually described with a higher-order Poincaré (HOP) sphere, in which the basis states are more general orthogonal states that incorporate both SAM and OAM[12,54]. In our case, a TAM beam can be described as follows

$$\text{TAM} = (O_x + e^{-i\varphi} \cdot O_y)/\sqrt{2}, \tag{2}$$

in which

$$O_x = (\text{LP}_{11a-x} + e^{-i\theta_1} \cdot \text{LP}_{11b-x})/\sqrt{2}, \tag{2-a}$$

$$O_y = (\text{LP}_{11a-y} + e^{-i\theta_2} \cdot \text{LP}_{11b-y})/\sqrt{2}. \tag{2-b}$$

Therefore, a TAM beam can be synthesized with four LP$_{11}$ modes (i.e., LP$_{11a-x}$, LP$_{11a-y}$, LP$_{11b-x}$, LP$_{11b-y}$) and carries three phase terms (i.e., $\theta_1$, $\theta_2$, and $\varphi$). More details about these phase terms are explained below.

(1) $\theta_1$ is the phase difference between the LP$_{11a-x}$ and LP$_{11b-x}$ modes, determining the phase structure of the E$_x$ component of the TAM beam.

(2) $\theta_2$ is the phase difference between the LP$_{11a-y}$ and LP$_{11b-y}$ modes, determining the phase structure of the E$_y$ component of the TAM beam.

(3) $\varphi$ is the phase difference of the terms O$_x$ and O$_y$, determining the polarization state of the TAM beam.

As an example, Fig. 5 shows some of the synthesized TAMs, showing their intensity/polarization patterns, and the phase structures of their E$_x$ and E$_y$ components. In these examples, the phase terms $\theta_1$ and $\theta_2$ are chosen to be ±π/2 so that both E$_x$ and E$_y$ components carry OAM$_{\pm1}$; while $\varphi$ is appropriately chosen to produce the linear polarization (e.g., [$\theta_1$, $\theta_2$, $\varphi$]=[π/2, π/2, π/2] or [-π/2, -π/2, -π/2]) or the vortex-like polarization (e.g., [$\theta_1$, $\theta_2$, $\varphi$]=[π/2, π/2, 0] or [-π/2, π/2, π/2]), respectively. For the structured light with [$\theta_1$, $\theta_2$, $\varphi$]=[π/2, π/2, 0] considered as an example, it has clockwise vortex-like polarization distribution, and the components of $E_x$ and $E_y$ carry $OAM_{+1}$, $OAM_{-1}$, respectively. By appropriately choosing the phase differences [$\theta_1$, $\theta_2$, $\varphi$], one can traverse the entire HOP sphere and synthesize the desired light beams carrying $OAM_{\pm1\text{-}x}$, $OAM_{\pm1\text{-}y}$, and SAM[41]. In order to generate the aforementioned TAM beams, definitely the powers and phases of these four $LP_{11}$ modes should be accuracy adjusted and one should minimize the thermal crosstalk on the SOI chip.

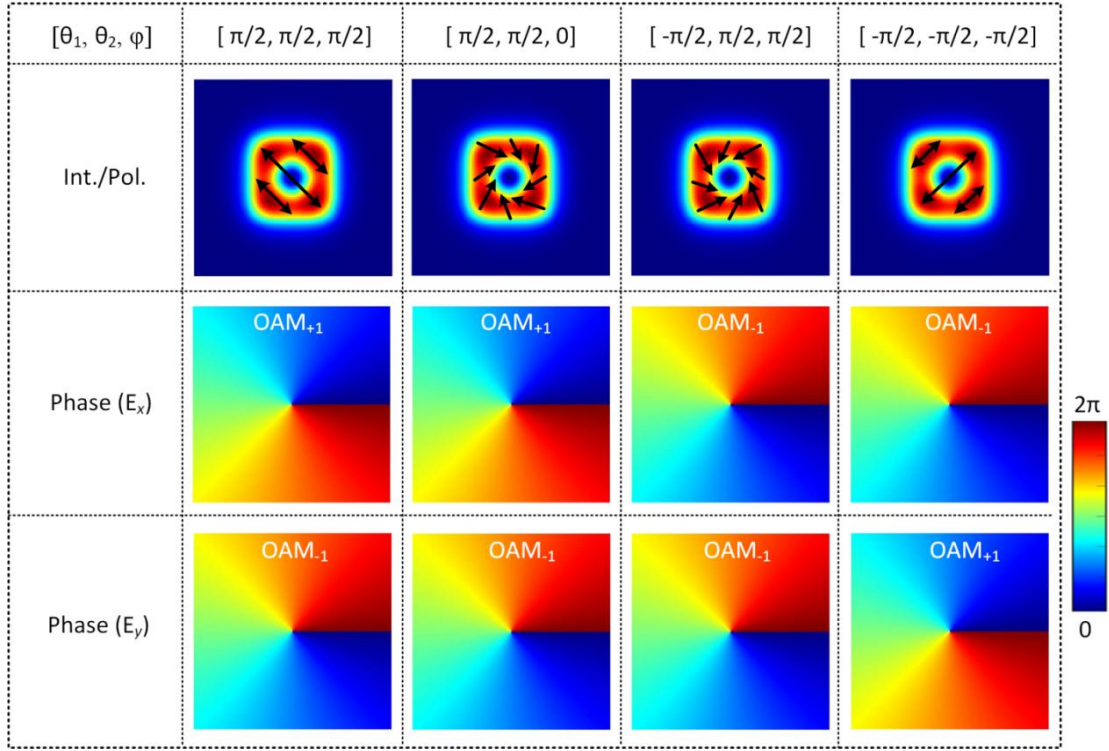

**Fig. 5  The synthesis of the TAM beams.** The simulated intensity/polarization patterns, phase structure of the $E_x$ component, and phase structure of the $E_y$ component of the synthesized TAM beam by stimulating $LP_{11a\text{-}x}$, $LP_{11a\text{-}y}$, $LP_{11b\text{-}x}$, $LP_{11b\text{-}y}$ modes with different phases [$\theta_1$, $\theta_2$, $\varphi$]. When [$\theta_1$, $\theta_2$, $\varphi$]=[π/2, π/2, π/2], the structured light has a donut intensity, 135° linear polarization and carries $OAM_{+1\text{-}x}$ and $OAM_{+1\text{-}y}$; When [$\theta_1$, $\theta_2$, $\varphi$]=[π/2, π/2, 0], the structured light has a donut intensity, clockwise-vortex polarization and carries $OAM_{+1\text{-}x}$ and $OAM_{+1\text{-}y}$; When [$\theta_1$, $\theta_2$, $\varphi$]=[-π/2, π/2, π/2], the structured light has a donut intensity, anticlockwise-vortex polarization, and carries $OAM_{-1\text{-}x}$ and $OAM_{-1\text{-}y}$; When [$\theta_1$, $\theta_2$, $\varphi$]=[-π/2, -π/2, -π/2], the structured light has a donut intensity, 45° linear polarization, and carries $OAM_{-1\text{-}x}$ and $OAM_{+1\text{-}y}$.

## Discussion

In summary, we have demonstrated an all-on-chip reconfigurable structured light generator by incorporating an SOI chip with a silica chip.  Our work advances the nascent field of PICs for structured light creation, showing how to both create and control the light all on-chip without the clumsy external free-space conversion (which negates the very benefit of starting on-chip in the first place). We demonstrate its power by going beyond the state-of-the-art and showing full angular momentum control, from scalar orbital angular momentum (OAM) to vectorial combinations, made possible by full polarisation control. Not only do we create this on-chip, but we also keep it there, with direct control and delivery through a waveguide. Further, our device has a fibre input and fibre output for a truly integrated, compact and reconfigurable solution, showing excellent performance in modal spectrum, wavelength spectrum and speed. We believe that this makes our solution idea for applications such as mode division multiplexing in fibre, fibre sensing with structured light and on-chip quantum technologies based on the spatial modes of light.

The proposed structured light generator can synthesize more complex total angular momentum beams and it can be extended to generate higher-order structured light beams by increasing the modal capacity of the silica mode multiplexer and the output number of the SOI chip. Compared with the methods reported previously on-chip generating structured light, the scheme proposed here shows a prominent advantage of versatility, broad bandwidth, and high conversion efficiency. Since the silica MBW can be butt-coupled efficiently to an FMF with a low coupling loss of <2 dB or less for all six LP modes by using the proposed multimode segmented waveguide [53], the structured light beams synthesized by the present all-on-chip generator can be guided into an OAM fibre conveniently for remote applications. Further improvement should be attention to further minimizing the excess loss, which is possible by introducing a variable power splitter [55] as well as adopting a compact architecture.

## Materials and Methods

1. **Silicon photonic chip design**

Figure. S1(a) shows the schematic configuration of the SOI chip that produces three pairs of $TE_0/TM_0$ modes whose power ratios and phase shifts can be tuned thermally. The SOI chip consists of a PBS, seven 1×2 MMI 3-dB couplers, six variable optical attenuators (VOAs; $A_1$, $A_2$, …, $A_6$), six phase shifters (PSs; $P_1$, $P_2$, …, $P_6$), three PSRs ($R_1$, $R_2$, $R_3$), and four ECs ($EC_i$,

EC$_1$, EC$_2$, EC$_3$). Fig. S1(b) shows the cross-section of the silicon photonic waveguide, which has a 220-nm-thick silicon core layer surrounded by silica, and a metal micro-heater is located on the top with a separation of $h_g$=1.5 μm to balance the heating efficiency and optical absorption loss. The PBS used here is designed based on the bent directional coupler as shown in Fig. S1(c) proposed in[49]. The PSR used here works on the principle of mode hybridness of the tapered ridge waveguide, and a directional coupler is further used to separate the TE$_0$ and TM$_0$ modes, as shown in Fig. S1(d)[51,52]. The 1×2 3-dB coupler is designed based on MMI as shown in Fig. S1(e). The VOAs are designed based on a symmetric 1×1 Mach-Zender interferometer (MZI) as shown in Fig. S1(f). The designed phase shifters are shown in Fig. S1(g), which utilizes the effective thermo-optical coefficient difference between the TE$_0$ and TM$_0$ modes in a silicon photonic waveguide, and the waveguide width of 0.45 μm is chosen to balance the thermal efficiency and transmission loss[50]. The EC based on an inverse taper structure at the two ends of the SOI is used for the butt coupling between the SOI chip and HSMF or silica single mode waveguide, as shown in Fig. S1(h). The inverse taper is linearly tapered from 0.45 to 0.16 μm, thus well matching the mode field of the HSMF or silica single mode waveguide. The key parameters of the designed silicon components are summarized in Table S1.

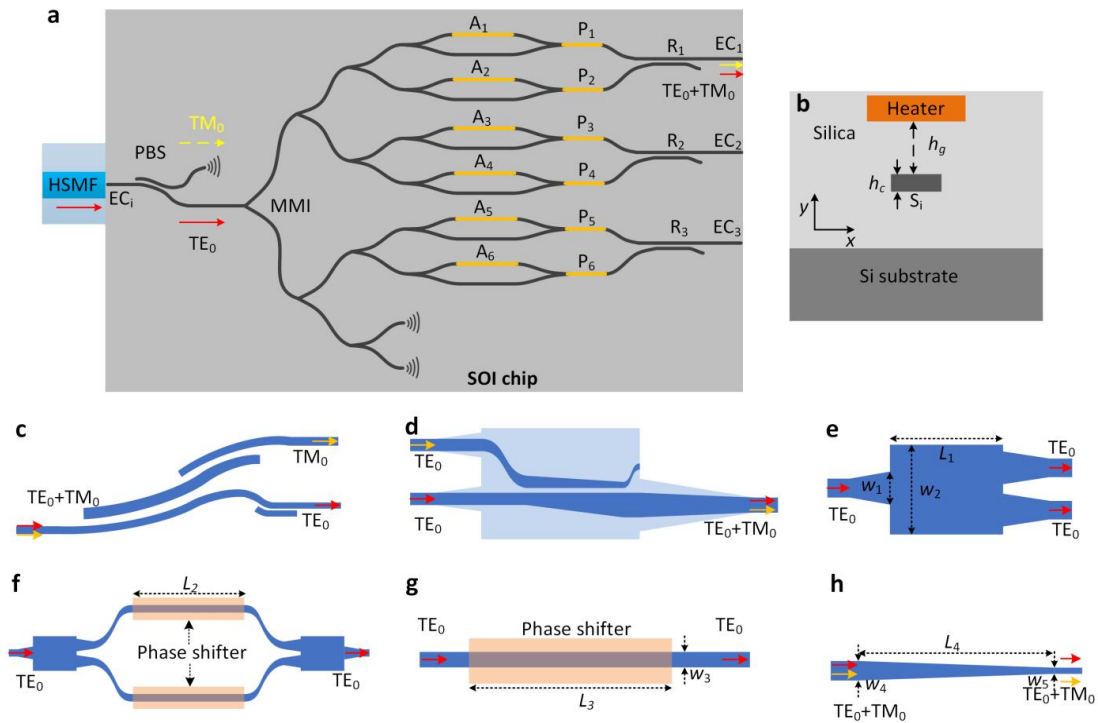

**Figure S1. Silicon photonic chip.** (a) Schematic configuration of the SOI chip, which consists of a polarization beam splitter (PBS), seven 1×2 3-dB multimode interference (MMI) couplers, six

variable optical attenuators (VOAs; $A_1$, $A_2$, …, $A_6$), six phase shifters (PSs; $P_1$, $P_2$, …, $P_6$), three polarization-splitter rotators (PSRs, $R_1$, $R_2$, $R_3$), and four edge couplers (EC; $EC_i$, $EC_1$, $EC_2$, $EC_3$). **b** The silicon waveguide cross-section. The schematic configuration of the (**c**) PBS, (**d**) PSR, (**e**) 1×2 MMI coupler, (**f**) VOA, (**g**) PS, and (**h**) EC.

**TABLE S1.** The key parameters of the designed silicon photonic devices.

| Parameters | $w_1$ | $w_2$ | $w_3$ | $w_4$ | $w_5$ | $L_1$ | $L_2$ | $L_3$ | $L_4$ |
|---|---|---|---|---|---|---|---|---|---|
| Value (μm) | 1.6 | 4 | 0.45 | 0.45 | 0.16 | 13.75 | 50 | 50 | 180 |

2. **Silica mode multiplexer design**

The three-channel silica mode multiplexer consists of two cascaded adiabatic directional couplers (ADC #1, ADC #2), a mode rotator and has a total length L=6000 μm as shown in Fig. S2(a). The silica mode multiplexer is polarization-insensitive due to the low birefringence and low index contrast. The inset of Fig. S2(a) shows the cross-section of the silica waveguide, whose index contrast is about 1.5%. In particular, the silica waveguides are designed with two different heights, i.e., $h_1$=6.5 μm and $h_2$=4, to realize efficient conversion of the $LP_{01}$ and $LP_{11b}$ modes which have different mode-field symmetry[56]. The core sizes of the single mode input waveguide and multimode output waveguide are 4.0 × 4.0 μm² and 6.5 × 6.5 μm², respectively. At the output port, the silica multimode bus waveguide (MBW) can couple to the FMF with high efficiency and low crosstalk for the applications. The operation principle of the mode multiplexer is shown with details in Fig. S2(b). Here the $LP_{01}$ mode launched into Pt #1 couples to the $LP_{11a}$ mode of the MBW via ADC #2; the $LP_{01}$ mode launched into Pt #2 passes through the two ADCs directly; meanwhile, the $LP_{01}$ mode launched into Pt #3 couples to the $LP_{11a}$ mode of the MBW via ADC #1 first and then rotates to $LP_{11b}$ mode with a mode rotator based on the tilt-etched dual-layer waveguide. The simulated transmissions for the lights launched into Pt #2, Pt #3, and Pt #1 are shown in Fig. S2(b, c, d), respectively, while the insets show the corresponding light propagation for the operation at the wavelength of 1550 nm. The simulation results show the designed mode multiplex has an ultra-low loss less than 0.04 dB and crosstalk less than -32 dB for the $LP_{01}$, $LP_{11a}$, and $LP_{11b}$ modes in the wavelength range of 1500-1600 nm. The ultra-low loss and low crosstalk can be attributed to the principle of adiabatic mode evolution in the designed ADCs.

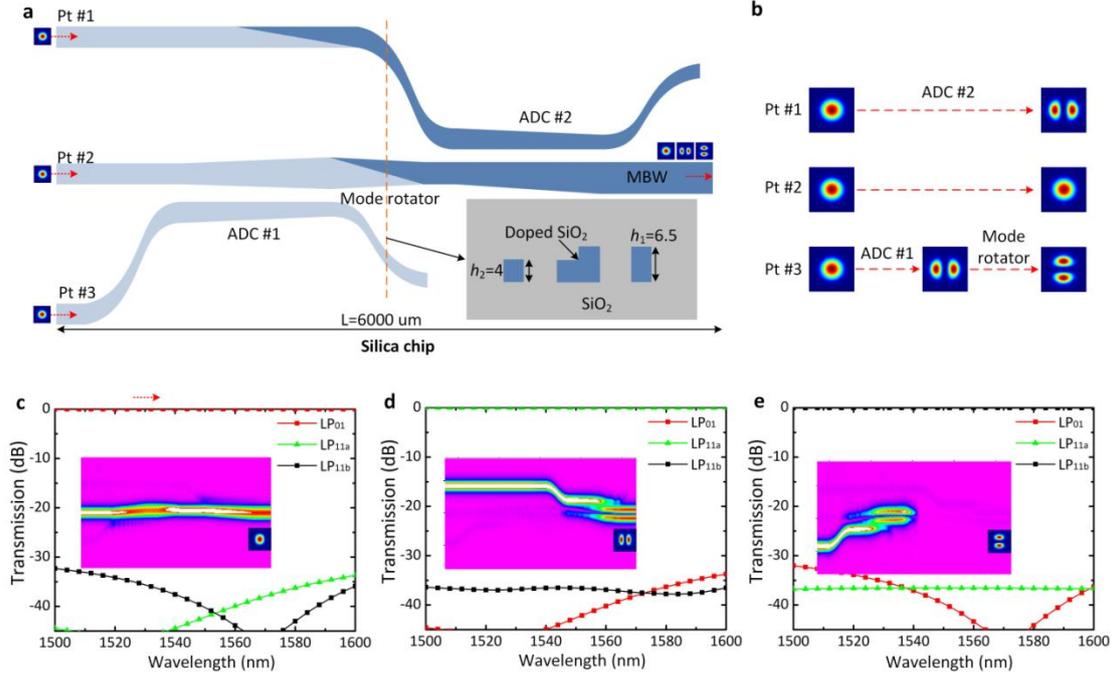

**Figure S2. Silica chip.** (**a**) Schematic configuration of the silica mode multiplexer, consisting of three input ports (Pt #1, Pt #2, Pt #3), two adiabatic directional couplers (ADC #1 and ADC #2), a mode rotator, and an output silica multimode bus waveguide (MBW). The inset shows the cross-section of the silica waveguide, it has a doped silica core and a pure silica cladding, while the silica single mode waveguide and multimode waveguide are designed with the core heights of $h_2$=4 μm and $h_1$=6.5 μm, respectively. (**b**) Operation principle: the $LP_{01}$ mode launched into Pt #1 couples to the $LP_{11a}$ mode of the MBW via ADC #2; the $LP_{01}$ mode launched into Pt #2 passes through the two ADCs directly; the $LP_{01}$ mode launched into Pt #3 is coupled to the $LP_{11a}$ mode of the MBW via ADC #1 first and is then rotated to the $LP_{11b}$ mode with a mode rotator based on the tilt-etched dual-layer waveguide. Simulated transmissions when light launched into (**c**) Pt #2, (**d**) Pt #1, and (**e**) Pt #3, the insets show the corresponding light propagation at 1550 nm.

## 3. Transmission matrix

The transmission of the structured light generator can be depicted by the following transmission matrix:

$$E_{out} = C_P C_A E_{in}, \qquad (3)$$

in which $E_{in}$ is the excited six LP modes basis sets, $C_A$ is the amplitude matrix controlled by the six VOAs, and $C_p$ is the phase matrix controlled by the six PSs. They are given as:

$$E_{in} = [LP_{11a-x} \quad LP_{11a-y} \quad LP_{01-x} \quad LP_{01-y} \quad LP_{11b-x} \quad LP_{11b-y}]^T, \qquad \text{(3-a)}$$

$$C_A = \begin{bmatrix} A_1 & 0 & 0 & 0 & 0 & 0 \\ 0 & A_2 & 0 & 0 & 0 & 0 \\ 0 & 0 & A_3 & 0 & 0 & 0 \\ 0 & 0 & 0 & A_4 & 0 & 0 \\ 0 & 0 & 0 & 0 & A_5 & 0 \\ 0 & 0 & 0 & 0 & 0 & A_6 \end{bmatrix} \tag{3-b}$$

$$C_p = \begin{bmatrix} e^{-i\varphi_1} & e^{-i\varphi_2} & e^{-i\varphi_3} & e^{-i\varphi_4} & e^{-i\varphi_5} & e^{-i\varphi_6} \end{bmatrix} \tag{3-c}$$

## 4. Fabrication

The SOI wafer used here has a 3-μm-thick buffer layer and a 220-nm-thick top-silicon core layer. The processes of electron beam lithography (EBL) and inductively coupled plasma (ICP) were used to form the bi-level ridge waveguides with a slab thickness of 70 nm. A 2.3-μm-thick silica upper cladding was deposited with the plasma-enhanced chemical vapor deposition (PECVD) process, and the 300-nm Cr/Ti metal layer was embedded into the cladding as the heater.

The silica chip was fabricated with a silica wafer with a silicon substrate, a 10-μm-thick silica buffer layer, and a 6.5-μm-thick doped silica-core layer. Here the index contrast of silica waveguides is about 1.5%. The silica core layer was etched with the Cr mask by using the inductively coupled plasma (ICP) process. Finally, a 15-μm silica upper cladding was formed by using the flame hydrolysis deposition (FHD) technology.

## 5. Measurement

Figure S3(a) shows the experimental setup for mode field detection. Here a tunable laser (TL) @1550 nm and a fibre polarization controller (PC) connected with an HSMF were used at the input side to be butt-coupled efficiently to the SOI chip, while the SOI chip was then butt-coupled to the silica chip. Finally, the light output from the silica MBW was collimated by a 20× objective, then passed through a QWP (with a horizontal fast axis), a polarizer, and was finally captured with a CCD camera. A multichannel voltage source (MVS) was used to power the six VOAs and six PSs, so that the power ratios and the phase shifts of these six $TE_0/TM_0$ channels can be controlled. Fig. S3(b) shows the experimental setup for measuring the transmission of generated six LP-mode basis sets. Here amplified spontaneous emission (ASE) was used as a light source, and the output LP modes were received with a few-mode fibre (FMF) and sent into an optical spectrum analyzer (OSA).

The interference setup for detecting the synthesized OAM light beam is shown in Fig. S3(c). Here light from a tunable laser was divided into two parts with a power ratio of 50%:50% through a fibre 3-dB coupler. One part was sent into the SOI chip and used as the signal beam, and the other one was used as the reference Gaussian beam for interference. The signal light passed through the chips and finally was expanded with a 20× Objective, while the reference Gaussian beam was collimated with a fibre collimator. When the signal light and the reference Gaussian beam were combined with a nonpolarizing beam splitter (NPBS), and the generated interference pattern was then captured with a CCD camera. The power ratio of these two routes was balanced with two optical attenuators, and their polarization states were calibrated with two fibre PCs.

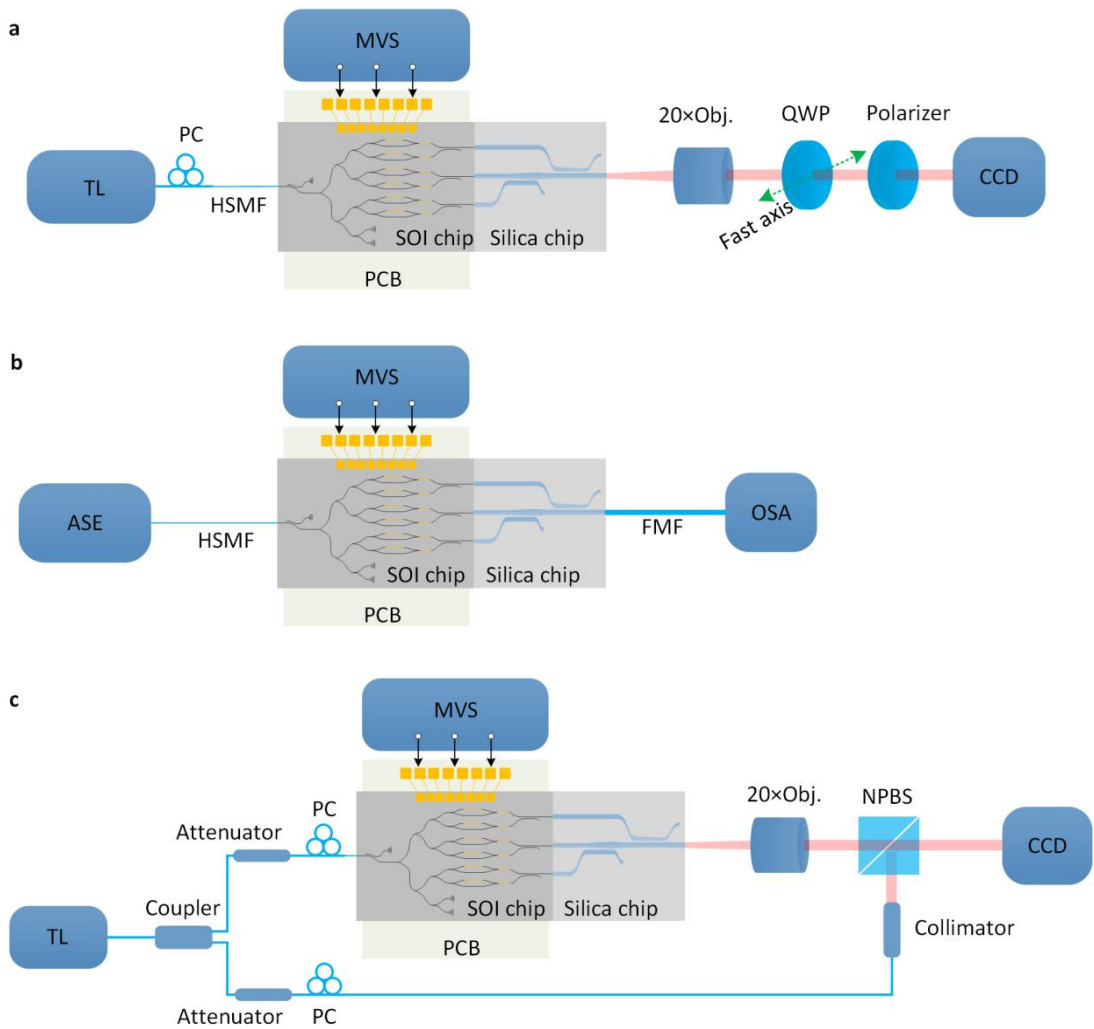

**Figure S3. Measurement experiment setup.** (**a**) Experiment setup for mode field detection, (**b**) Experiment setup for the measurement of mode transmissions. (**c**) The experiment setup for detecting the OAM beams. TL: tunable laser, ASE: amplified spontaneous emission, MVS: multichannel voltage souce, PCB: printed circuit board, HSMF: high-NA single mode fibre, PC: polarization controller, MVS: multi-channel voltage source, FMF: few-mode fibre, OSA: optical

spectrum analyzer, QWP: quarter-wave plate, NPBS: nonpolarizing beam splitter, CCD: charge coupled device.

## 6. Butt-coupling loss and PBS, PSR performances

The HSMF and SOI waveguide coupling was achieved with a silicon EC based on an inverse taper waveguide. The measured coupling loss for each HSMF-SOI facet is shown in Fig. S4(a), and it is 2 dB for the $TE_0$ mode in the 1520-1600 nm wavelength range. The coupling between the SOI waveguide and a 4×4 $\mu m^2$ single mode silica waveguide is also achieved with the same EC. Fig. S4(b) shows the measured SOI-Silica butt-coupling loss, which is about 0.4/1.5 dB for the $TE_0/TM_0$ modes in the wavelength range around 1560 nm.

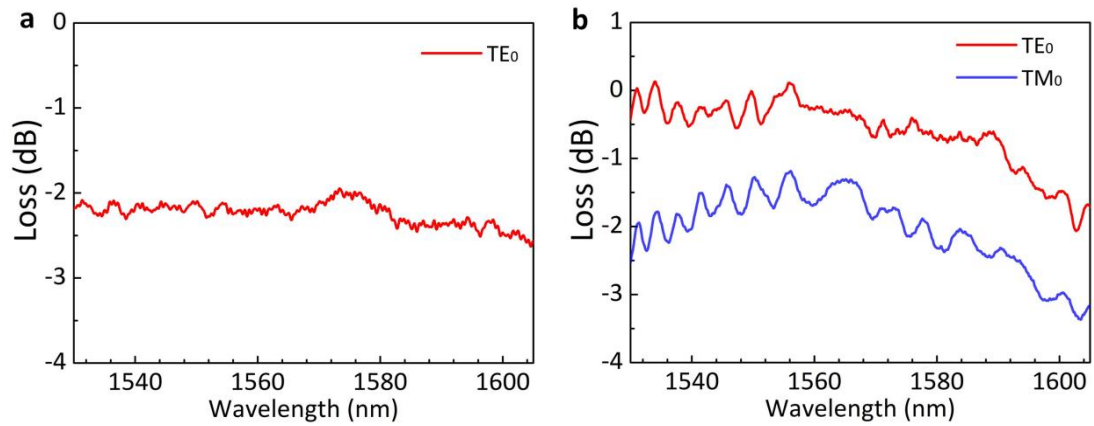

**Figure S4.** (**a**) The measured butt-coupling loss for the $TE_0$ mode between an HSMF and a silicon waveguide; (**b**) The measured butt-coupling losses for $TE_0/TM_0$ modes between a silicon waveguide and a silica waveguide.

The measure results for the testing silicon PBS and PSR fabricated on the same chip are shown in Fig. S5(a, b), respectively. The PBS has a low loss of <1 dB and low crosstalk <-15 dB for both the $TE_0$ and $TM_0$ modes. The PSR has a low loss of <1 dB and low crosstalk <-15 dB for both the $TE_0$ and $TM_0$ modes.

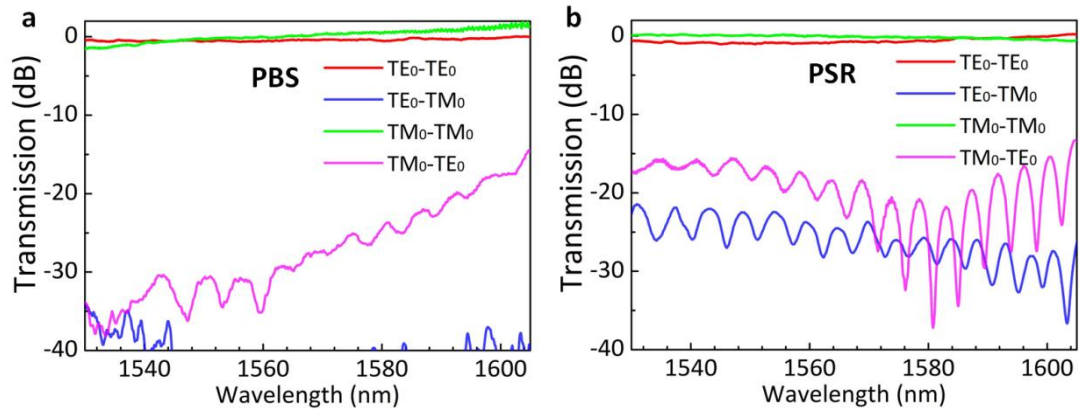

**Figure S5.** The measured transmissions for (**a**) the PBS, (**b**) the PSR.

## Data availability statement

All data are available in the main text or the supplementary materials.

## Conflict of interest

All other authors declare they have no competing interests.

## Acknowledgments

This work is supported by National Natural Science Foundation of China (NSFC) (62375238, 92150302, U23B2047, and 62321166651), Zhejiang Provincial Major Research and Development Program under Grant 2021C01199, The Fundamental Research Funds for the Central Universities, The Leading Innovative and Entrepreneur Team Introduction Program of Zhejiang (2021R01001).

## Author contributions

W. K. Z. and D. X. D. conceived the idea. W. K. Z and X. L. Y. contributed equally to this work. W. K. Z and R. R. L. designed and fabricated the silicon chip. X. L. Y. designed and fabricated the silica chip. K. Z and X. L. Y. built the setup and carried out the experiment. All authors discussed the results and contributed to the manuscript. W. K. Z, X. L. Y., D. X. D. and A. F. wrote the manuscript. D. X. D. supervised the project.